\documentclass{aa}  % [referee]{aa} for a referee version  

\usepackage{graphicx}
\usepackage{txfonts}
\usepackage{hyperref}
\usepackage{bm}
\bibpunct{(}{)}{;}{a}{}{,} % to follow the A&A style

\newcommand{\zphoto}{z_\mathrm{photo}}
\newcommand{\Pzphoto}{P(z_\mathrm{photo})}
\newcommand{\SPzphoto}{\Sigma(P(z_\mathrm{photo}))}

\newcommand{\zspec}{z_\mathrm{spec}}
\newcommand{\Szspec}{\Sigma(z_\mathrm{spec})}
\newcommand{\msestar}{\mathrm{MSE}^\ast}

\usepackage{booktabs}
\usepackage{multirow}
\newcommand{\bfs}{\boldmath\bfseries}   % spójne pogrubianie w tabelach (przełącznik na komórkę)

\begin{document}

\title{Quasar photometric redshifts beyond the spectroscopic coverage}
\subtitle{Uncertainty models and redshift distributions in the Kilo-Degree Survey DR5}

\titlerunning{Quasar photo-$z$ beyond the spectroscopic coverage}
\authorrunning{Kacper Drabicki et al.}

\author{
    Kacper Drabicki\inst{1}
    \and Szymon J. Nakoneczny\inst{1}\thanks{Corresponding author: S. J. Nakoneczny nakoneczny@cft.edu.pl}
    \and Maciej Bilicki\inst{1}
}

\institute{
    Center for Theoretical Physics, Polish Academy of Sciences, Al. Lotników 32/46, 02-668 Warsaw, Poland\\
}

\date{}

\abstract
% Context
{Photometric redshifts (photo-$z$) and their distributions underpin cosmology with photometric quasars, tracers in angular clustering and cross-correlations. Further progress requires trustworthy uncertainties, especially where data reach beyond the spectroscopic training set.}
% Aims
{We compare how machine-learning frameworks estimate quasar photo-$z$ uncertainties and how they reconstruct the redshift distribution $n(z)$ under controlled data-quality shifts, to establish which model to prefer as a first step towards calibrated tomography.}
% Methods
{Using Kilo-Degree Survey DR5 photometry and DESI DR1 spectroscopic quasars, we train artificial neural networks (ANNs), Mixture Density Networks (MDNs) and Bayesian Neural Networks (BNNs) with Gaussian-mixture outputs, and add a self-organizing map (SOM) as a direct $n(z)$ estimator. We evaluate them on four subsets, with and without magnitude extrapolation and missing bands, through the negative log-likelihood, the probability integral transform, point-estimate accuracy, and the bias of the binned $n(z)$ moments; degeneracies are sought by clustering the predicted PDFs.}
% Results
{At least two mixture components are essential: a single Gaussian is miscalibrated and produces the most catastrophic point-estimate outliers, more than the ANN. No model performs best everywhere. On well-covered data the five-component MDN, three-component BNN and SOM reconstruct $n(z)$ almost perfectly; under faint extrapolation the BNN gives the best likelihoods, while in the hardest faint-plus-missing regime the single Gaussian becomes the best-calibrated. Regarding point estimates and tomographic binning, the uncertainty models perform better than the ANN, while the SOM fails out-of-distribution. PDF clustering exposes distinct colour--redshift degeneracies expected to proliferate for fainter samples.}
% Conclusions
{The best model is regime- and application-dependent: multi-component MDNs or BNNs are needed for clean binning and are the only intrinsically calibrated choice on a well-covered golden sample, whereas the single Gaussian does the opposite -- miscalibrated on the easiest data, yet best-calibrated and robust under the strongest shift. This is a first step towards a full, calibrated comparison of quasar photo-$z$ pipelines.}


\maketitle
\nolinenumbers

\section{Introduction}
\label{sec:intro}

Photometric redshift estimates (photo-$z$) have become a cornerstone of modern cosmological analyses. Although they are less precise than spectroscopic measurements, photometric techniques enable redshift estimation for orders of magnitude more objects, making them indispensable for large-area surveys and statistical cosmology \citep[e.g.][]{Hildebrandt_2010,Salvato_2019,Newman_2022}. Photo-$z$ are commonly used to construct tomographic redshift bins, which are used for cosmological inference from large-scale structure probes such as weak gravitational lensing and galaxy clustering. These distributions are often calibrated using additional data.

Cosmological analyses rely not only on point estimates of redshift but on a robust characterization of the associated uncertainties, since biases in the reconstructed redshift distribution propagate directly into systematic errors in observables that depend on accurate distances. In analyses involving the lensing of background sources or of the cosmic microwave background, such errors bias the lensing efficiency kernels entering the cross-correlation signal, and can lead to incorrect constraints on parameters such as $\sigma_8$ and $\Omega_\mathrm{m}$; for point-like sources such as quasars, this concerns their role as foreground tracers rather than as shape-measured lensing sources. Similarly, in galaxy clustering measurements, biases in the redshift distribution affect inferred comoving distances and correlation functions, impacting both the amplitude and scale of the signal. In this paper we focus on both individual redshift uncertainties and the resulting redshift distributions, which we reconstruct from the per-object photo-$z$ PDFs.

In case of machine learning (ML), the photo-$z$ uncertainties are commonly decomposed into aleatoric and epistemic components. Aleatoric uncertainty arises from intrinsic degeneracies in the data, where different redshift values correspond to similar photometric colours, even in the presence of abundant training data. Epistemic uncertainty, on the other hand, is driven by incomplete coverage of the feature space by the training set, reflecting the model’s lack of knowledge in poorly sampled regions.

Aleatoric uncertainty in photo-$z$ estimation is often modelled using Mixture Density Networks (MDNs), i.e. artificial neural networks (ANNs) that parameterize the output as a Gaussian Mixture Model (GMM) \citep{DIsanto_2018,Zhang_2024,Chen_2025}. This approach allows the prediction of full probability density functions (PDFs) rather than single-point estimates. An alternative strategy frames redshift estimation as a classification problem over discrete redshift bins \citep{Pasquet_2019}. While this method avoids explicit assumptions about the functional form of the PDF, its performance and resolution are inherently limited by the chosen bin width.

To capture epistemic uncertainty, Bayesian Neural Networks (BNNs) are commonly employed, in which the network weights are treated as random variables with prior distributions, typically Gaussian \citep{Zhou_2022,Jones_2024}. A computationally efficient approximation to full Bayesian inference is Monte Carlo (MC) dropout \citep{Luo_2024}, in which a subset of weights is randomly deactivated not only during training but also at inference, providing a variational approximation to a BNN.

Here we focus on the Kilo-Degree Survey (KiDS) Data Release 5 (DR5). Within the context of KiDS, photometric redshifts (photo-$z$) have been estimated for both quasars and galaxies \citep{Bilicki_2021,Hildebrandt_2021,Nakoneczny_2021,Li_2022,Vakili_2023,William_2025_a,William_2025_b}. However, a comprehensive characterization of the associated uncertainties has not yet been established.

In this work we focus exclusively on quasars. As tracers of the large-scale structure they give access to much higher redshifts than regular galaxies. The most powerful cosmological constraints from quasars come from spectroscopic samples such as the Sloan Digital Sky Survey (SDSS) and the Dark Energy Spectroscopic Instrument (DESI); photometrically selected quasars nonetheless remain valuable, being far more numerous and complete than spectroscopic samples. Recent analyses with photometric quasars constrain their host halo masses \citep[e.g.][]{Luo_2024,Eltvedt_2024,William_2025_b}, in particular by dividing them into obscured and non-obscured \citep{Petter_2023}. Quasars, or more generally active galactic nuclei, constitute an important fraction of radio-detected sources, so their redshift distributions are key to extracting cosmological information from the largest radio samples, such as those from the Low-Frequency Array \citep[LOFAR, e.g.][]{Alonso_2021,Nakoneczny_2024}. In this context, we aim to investigate how photometric redshift uncertainties can be reliably estimated for quasars and how they affect the inferred redshift distributions.

Photometric redshift estimation for quasars has a long history predating deep-learning approaches. Empirical colour--redshift relations were established with early SDSS data \citep{Richards_2001,Weinstein_2004}, followed by machine-learning point-estimate methods \citep[e.g.][]{Brescia_2013}, while template-fitting methods tailored to AGN, accounting for variability and the relative contribution of nuclear and host-galaxy light, achieved accuracies competitive with galaxies in deep multiband fields \citep{Salvato_2009,Salvato_2011,Fotopoulou_2018}; see \citet{Salvato_2019} for a review. Probabilistic approaches producing full redshift PDFs for quasars were pioneered with extreme deconvolution applied to SDSS photometry \citep[XDQSOz;][]{Bovy_2011,Bovy_2012}, and have since included asymmetric-distribution regression \citep{Yang_2017} and hybrid template--ML frameworks with explicitly calibrated PDFs developed for radio-selected AGN \citep{Duncan_2018,Duncan_2021,Duncan_2022}. Most recently, all-sky quasar samples with photometric redshifts, such as Quaia \citep{StoreyFisher_2024}, have been used directly for cosmological cross-correlations. What remains missing in this literature is a controlled, like-for-like comparison of the two main uncertainty paradigms -- aleatoric modelling via mixture densities and epistemic modelling via Bayesian networks -- for quasars, in particular under the data-quality perturbations most relevant in practice: extrapolation beyond the training magnitude range and cases of missing photometric bands.

Cosmological analyses place requirements on the properties of photometric redshifts. Such requirements have been formulated for the weak-lensing source samples of Stage-IV surveys such as \textit{Euclid} \citep{Laureijs_2011} and LSST \citep{Mandelbaum_2018}. Photometric quasar samples are not used as weak-lensing sources, being point-like, but as tracers in angular clustering and in cross-correlations with CMB lensing; for these applications no comparable set of quantitative photo-$z$ requirements has been established, and the residual uncertainty of the redshift distribution is instead typically marginalized over via parametric forms of $n(z)$ \citep[e.g.][]{Myers_2007,Alonso_2023,Petter_2023,Nakoneczny_2024}. We assess our photometric redshifts in two complementary ways: the calibration of the per-object PDFs with the probability integral transform, and the accuracy of the reconstructed redshift distribution through the bias of its first two moments in redshift bins.

A widely used way to control redshift-distribution uncertainties in cosmological analyses is empirical calibration with self-organizing maps (SOMs), as in the cosmic-shear analyses of KiDS \citep{Wright_2020a, Hildebrandt_2021, Wright_2025b} and DES \citep{Myles_2021, Yin_2025}: the photometric space is projected onto a two-dimensional grid of cells, each calibrated with overlapping spectroscopy, and cells lacking adequate spectroscopic coverage are excised from the analysis, with the source sample restricted accordingly. For the flux-limited galaxy samples used in cosmic shear this is acceptable, because the spectroscopic calibration data cover the bulk of the source population. For quasars the situation differs in two respects. First, spectroscopic quasar samples are typically shallower than the photometric catalogues, so a large fraction of the latter lie in the faint regime where spectroscopic coverage is sparse or absent. Second, and independently of depth, quasar spectroscopy is rarely flux-limited in the bands used for the SOM: confirmation usually relies on optical spectra, whereas quasars also emit strongly in the UV, infrared, and radio, so spectroscopic samples are assembled from heterogeneous selections and need not be representative even where they are nominally deep enough.

Taken together, both effects mean that empirical per-cell calibration can be unreliable or unavailable over a substantial part of the relevant colour space. Discarding those regions would remove a major part -- in the faint regime, the majority -- of the very objects one aims to exploit, which is not an acceptable trade-off for quasar science. This motivates methods that instead extrapolate beyond the spectroscopic training distribution and provide reliable per-object uncertainties there, such as the MDN and BNN studied here. We therefore also compare a SOM-based reconstruction of $n(z)$ against our density-based estimates, to quantify where the empirical approach remains adequate and where it does not.

Producing calibrated tomographic redshift distributions from photometric data admits many distinct pipelines. The photometric redshift itself can be a point estimate or a full probability density; objects can be assigned to tomographic bins by that point estimate, by the density, or by their location in a self-organizing photometric map; and the redshift distribution within each bin can be built from the stacked densities, from the point estimates, or transferred from overlapping spectroscopy, with calibration applied to the per-object estimates (before binning) or to the binned $n(z)$ (after binning). These choices interact, and although individual combinations have been studied under specific assumptions \citep[e.g.][]{Hildebrandt_2010,Dahlen_2013,Buchs_2019,Schmidt_2020,Wright_2020a,Myles_2021,vandenBusch_2022}, a systematic like-for-like comparison across them is lacking \citep[see also][for a review]{Newman_2022}. The present work is a first step towards such a comparison: a controlled characterization of the photometric-redshift models themselves under quantified magnitude-extrapolation and missing-data shifts, without yet applying redshift calibration. We return to the full space of calibration options in Sect.~\ref{sec:discussion}, and defer a quantitative comparison of calibrated pipelines to future work.

The complex, degenerate spectral energy distributions of quasars make their photo-$z$ uncertainties considerably larger than those of galaxies, an ideal test case for the comparison above. Beyond comparing the uncertainty models, we test how well they recover the underlying spectroscopic redshift distribution, to establish which are most suitable for quasars in future analyses such as the Legacy Survey of Space and Time (LSST), and how the data should be selected or trimmed for a robust reconstruction.

To this end, we apply artificial neural networks (ANNs), MDNs, and BNNs to the problem of quasar photo-$z$ estimation in KiDS. We design multiple test sets to probe both aleatoric and epistemic uncertainties through controlled sampling strategies. Additionally, we use unsupervised learning with the predicted photo-$z$ probability density functions (PDFs) to investigate degeneracies present in the training data.

This paper is organized as follows. In Section~2 we describe the KiDS data set and the construction of the training and test samples. Section~3 presents the machine learning models and the evaluation metrics. The results are reported in Section~4 and interpreted in Section~5, and Section~6 provides the conclusions.

\section{Data}

\subsection{Kilo Degree Survey}

The Kilo-Degree Survey Data Release~5 \citep[KiDS-DR5,][]{Wright_2024} is the fifth and final public release of the ESO KiDS carried out with the VLT Survey Telescope and the OmegaCAM instrument. DR5 provides deep, homogeneous optical imaging over approximately 1347 deg$^{2}$ in the \textit{u, g, r, i} broadband filters, with two independent epochs in the $i$-band. The data reach a typical 5$\sigma$ limiting magnitude of $r\simeq24.8$ AB mag, with median seeing of about $0.7^{\prime\prime}$, enabling the detection of faint extragalactic sources and accurate shape measurements for weak-lensing analyses. When combined with near-infrared photometry from the VIKING survey \citep{Edge_2013}, KiDS-DR5 delivers multi-band \textit{ugriZYJHK$_s$} catalogues suitable for a wide range of cosmological and astrophysical applications \citep{Wright_2024}.

\subsection{Preprocessing}

We cross-match KiDS DR5 with quasars from the Dark Energy Spectroscopic Instrument \citep[DESI,][]{DESI_2016} Data Release 1 \citep[DR1,][]{DESI_2025} within 1 arcsec radius. We clean KiDS by applying the condition $\mathrm{MASK} < 2$. The DESI quasars were selected according to SPECTYPE 'QSO' and with $\mathrm{ZWARN} == 0$ to remove objects flagged with redshift warning. The data consists of 124638 quasars.

The DESI quasar sample used here as the spectroscopic reference has a specific selection function, based primarily on optical and WISE infrared photometry from the DESI Legacy Imaging Surveys \citep{Chaussidon_2023}, with a magnitude limit of $r < 23$. Photometric quasar catalogues trained on this spectroscopic sample will likely have a different selection function, both in depth and in candidate identification. The results of this paper therefore characterize model performance in the controlled regime where the training and test selections coincide (apart from the magnitude and missing-band perturbations introduced deliberately), and constitute an upper bound on the performance attainable on a real target catalogue.

\section{Methodology}

\subsection{Features}

Our initial feature space comprised GAaP magnitudes \citep{Kuijken_2008,Kuijken_2015,Kuijken_2019} in 10 photometric bands - $u$, $g$, $r$, $i1$, $i2$, $Z$, $Y$, $J$, $H$, and $K_s$ - which served as the primary inputs for quasar photo-$z$ estimation. We perform basic feature engineering by calculating magnitudes differences (colours). The resulting feature space comprises 55 dimensions. We also tested adding magnitude errors to the feature set; however, none of the experiments showed any improvement in performance.

\subsection{Test sets}
\label{sec:testsets}

\begin{figure}
    \centering
    \includegraphics[width=\hsize]{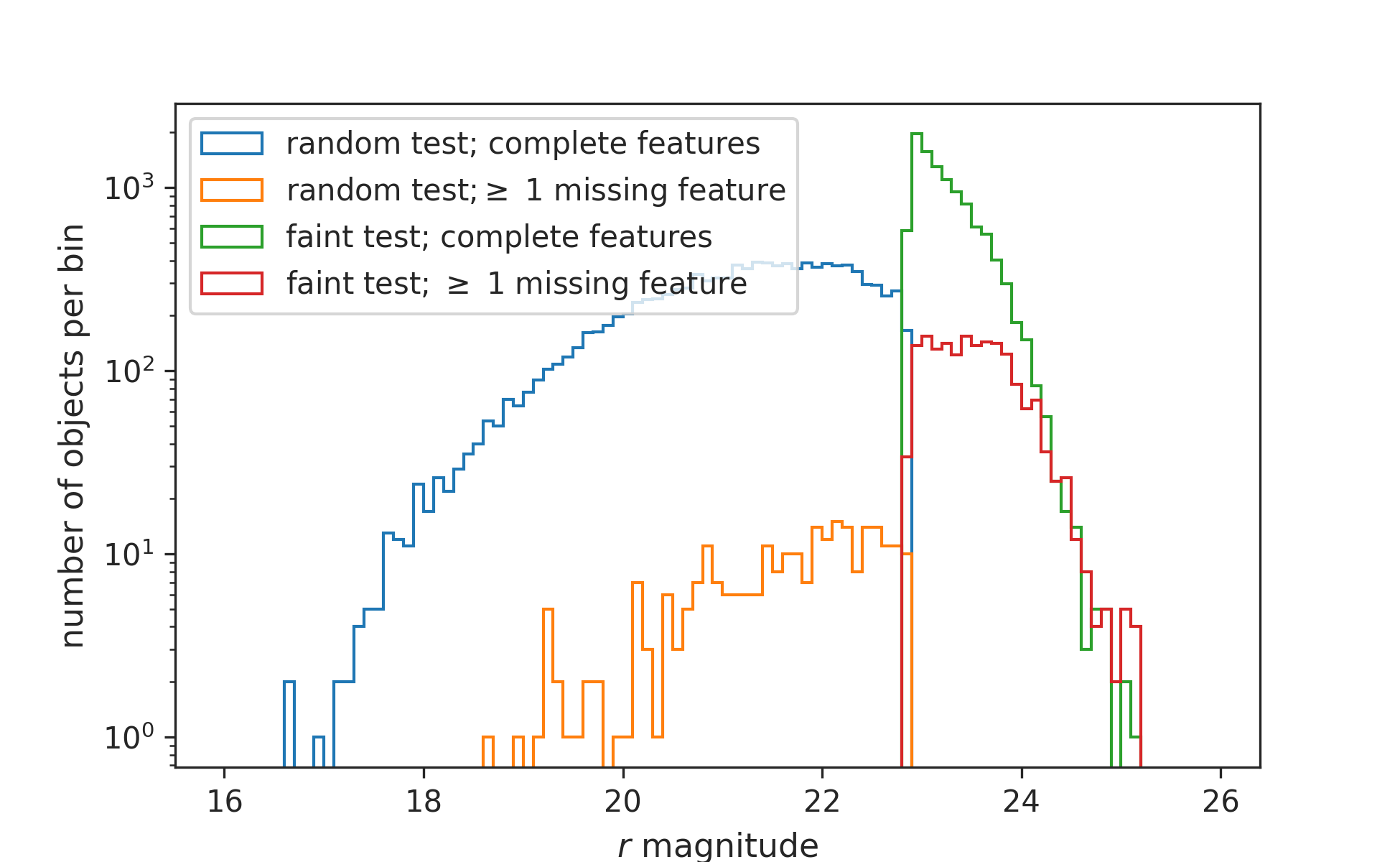}
    \caption{Distributions of test sets over $r$ magnitude. The faint-extrapolation test set includes the top 10\% of $r$ values, starting at $r > 22.9$. The random test contains a randomly selected 10\% of the objects with $r < 22.9$. Both test sets are comparable in size. The faint-extrapolation set appears more prominent only because its objects are concentrated in a much narrower $r$ range, whereas the random set is spread widely. We further split the both test sets into subsets with complete features and at least one missing feature.}
    \label{fig:r_mag}
\end{figure}

The overall sample has a median $r$-band magnitude of 21.5 and a 90th percentile of 22.9. To examine the extrapolation performance of the models, we defined a faint–extrapolation subset by selecting the top 10\% of objects ($r > 22.9$) with the faintest $r$-band magnitudes. After removing these objects from the main sample, we split the remaining data into training, validation, and test subsets in an 8:1:1 ratio. We hereafter refer to this test subset as the random test set.

Approximately 3.5\% of the objects lack at least one feature. Among these, the vast majority (88\%) are missing only a single magnitude. The most frequent gap is observed in the $u$ band, which accounts for 76\% of all incomplete cases, and is also one of the most important magnitudes for QSO photo-$z$ \citep{Nakoneczny_2021}.

We address these cases by replacing each missing entry with the maximum value of that feature in the training subset, hence, outside the faint–extrapolation subset; we use these same values to impute missing features in the faint–extrapolation subset for consistency. We then divide the random and the faint–extrapolation test sets into four subsets, named by the test set and the feature completeness and used with this shorthand throughout the paper: \emph{random/complete} and \emph{faint/complete} contain the objects with all features present in the random and the faint–extrapolation test, respectively, while \emph{random/missing} and \emph{faint/missing} contain those with at least one missing feature. Figure \ref{fig:r_mag} shows the distribution of $r$-band magnitudes in each of these four subsets. The four subsets contain, respectively, $11\,019$ (random/complete), $10\,702$ (faint/complete), $261$ (random/missing) and $1\,761$ (faint/missing) objects. The random-plus-missing subset is small -- bright sources rarely lack a band -- so its metrics are based on poor statistics and we treat them as indicative only.

A magnitude may be absent either because the source was not detected in a given band, or because that band was not observed at the source position, and the two cases are not distinguished in the data available to us. The over-representation of the $u$ band among the incomplete objects likely reflects, at least in part, the shallower depth of the $u$-band imaging, which affects all source classes in KiDS; for quasars, it may additionally be driven by Lyman-series absorption suppressing the $u$-band flux at $z \gtrsim 2.5$ \citep[e.g.][]{Fan_1999,Richards_2002}, in which case the pattern of missingness would itself carry redshift information. Magnitudes, and especially colours, are moreover ill behaved close to the detection limit; our imputation at the faint edge of the training distribution approximates such a limit, while keeping the treatment uniform across bands. Our experiments are by construction agnostic to the origin of the missing values: by removing measurements ourselves, we quantify the limiting impact of absent bands regardless of why they are absent in real data.

\subsection{Artificial neural networks}

We test three model families: a fully connected artificial neural network (ANN); Mixture Density Network (MDN) with Gaussian-mixture output layer (GMM); and a Bayesian Neural Network (BNN) with a GMM output layer. ANN estimates a single value of the photometric redshift $\zphoto$. In contrast, the MDN and BNN models estimate the parameters of a GMM distribution, which yields a probability distribution of the photometric redshift $\Pzphoto$. To enable a comparison between the ANN model and the MDN and BNN models, we obtain the point predictions $\zphoto$ from the MDN and BNN by taking the mean value of $\Pzphoto$. We also evaluated the median and mode; however, the mean yielded the best performance in terms of mean squared error.

For the MDN we vary the number of GMM components from one to six, and for the BNN from one to five. Throughout, we refer to these models as MDN~$N$ and BNN~$N$, where $N$ is the number of GMM components; for example, MDN~5 denotes the five-component MDN. All models use fully connected layers with non-linear activations. The ANN has 11 hidden layers of 256 neurons and outputs a single scalar. The MDNs use 12 hidden layers of 512 neurons; for multi-component MDNs, a dropout layer with rate 0.2 follows every two hidden layers to improve generalization. The MDNs predict the parameters of a GMM, returning a full PDF that captures aleatoric uncertainty. We adopt the mean of this distribution as the point prediction and the standard deviation as the associated uncertainty.

The BNN also predicts GMM parameters through 12 hidden layers of 512 neurons; however, the final three layers are replaced with Flipout Bayesian layers. These layers treat weights as probability distributions, enabling the model to capture both aleatoric uncertainty via the mixture output and epistemic uncertainty through the weight distributions. Because the BNN yields a different prediction at each evaluation, we generate 100 forward passes per object and take the final prediction as the average of their means. We quantify epistemic uncertainty as the variance of the 100 predicted means, and aleatoric uncertainty as the average of the predictive variances across the same draws; the total uncertainty is the square root of their sum.

\subsection{Model training}

For the ANN model, we adopt the mean squared error (MSE) as the loss function, calculated on the residuals $\Delta z = \zphoto - \zspec$, where $\zspec$ is the spectroscopic redshift. We train all the remaining models using the negative log-likelihood (NLL) of the predicted distribution evaluated at the probability value given by the true $\zspec$. We optimize all networks with Adam \citep{Kingma_2014}, a learning rate of $10^{-4}$, and early stopping after 30 epochs without improvement in the validation loss. To prevent data leakage, we fit all normalization parameters (mean and variance) exclusively on the training set. We train each model ten times and select, for the final evaluation, the realization that achieved the lowest loss on the validation set; the dispersion across these ten realizations is used throughout as the statistical scale of the metrics.

\subsection{Redshift distributions from a self-organizing map}
\label{sec:som_method}

As an independent, non-parametric alternative to the per-object density models, we also estimate the redshift distributions with a self-organizing map \citep[SOM;][]{Kohonen_1982}. SOMs have been used for photometric redshift estimation and object classification in astronomy \citep[e.g.][]{Geach_2012,Way_2012,CarrascoKind_2014}, while their use to map the multi-dimensional photometric space onto spectroscopic redshifts for the purpose of characterizing tomographic redshift distributions was pioneered by \citet{Masters_2015} and has since been applied in a number of cosmological analyses \citep[e.g.][]{Buchs_2019,Wright_2020a,Myles_2021,vandenBusch_2022,Wright_2025b,Yin_2025}. We stress that the SOM is used here only as a \emph{direct estimator} of $n(z)$, and not as a redshift-calibration scheme. We train a $20\times20$ SOM on the 55 photometric features of the training sample, with a neighbourhood width of $1.5$, a learning rate of $0.5$, and $1.5\times10^{6}$ iterations; the number of cells is chosen to be of order the square root of the training-sample size. We then assign each test object to its best-matching cell and estimate the redshift distribution of a test subsample from the spectroscopic redshifts of the training objects occupying the corresponding cells. In contrast to the KiDS cosmic-shear analyses \citep{Wright_2020a,Hildebrandt_2021}, where cells with insufficient spectroscopic coverage are removed and the source sample is restricted accordingly, we deliberately retain all objects and all cells: for a quasar sample, in which a large fraction of objects lie in the faint regime poorly covered by spectroscopy, discarding such cells would remove much of the sample of interest and defeat the purpose of the comparison. For a test set drawn from the same distribution as the training data, poorly covered cells carry few objects and contribute little to the aggregate metrics; cell removal matters far more in the transition to an independent inference catalogue. Retaining all objects therefore does not bias our comparison, while keeping it representative of how the method would have to be applied to a real quasar catalogue.

\subsection{Model evaluation}

To evaluate the point estimates (the mean of the predicted PDF) against the ANN, we use the redshift error
\begin{equation}
\label{eqn:z_error}
    \delta z = \frac{z_{\rm photo} - z_{\rm spec}}{1 + z_{\rm spec}},
\end{equation}
reporting its mean (bias) and standard deviation (scatter), together with the outlier fraction $\eta$, defined as the fraction of objects with $|\delta z| > 0.15$. Since the ANN serves only as a baseline and the primary goal of this work is to assess uncertainty estimation, we adopt as our main evaluation metric the NLL of the predicted redshift distribution evaluated at $\zspec$, averaged over the different test sets. Unlike these point-estimate metrics, the NLL evaluates the full predicted probability density function and therefore provides a direct measure of uncertainty quality.

To compare each model's total photo-$z$ distribution with the spectroscopic one, we build the spectroscopic distribution as a simple histogram of the reference redshifts $\Szspec$, while for the ML models we propagate the per-object PDFs by Monte Carlo sampling: for each object we draw a single random redshift from its predicted PDF and histogram the result, repeating this 5000 times and averaging the counts in each bin to suppress Monte Carlo fluctuations. We adopt a bin width of $\delta z = 0.3$ to focus on overall trends, and denote the resulting distribution $\SPzphoto$.

To assess the calibration of the predicted per-object PDFs, we use the probability integral transform \citep[PIT; e.g.][]{Dahlen_2013,Polsterer_2016,DIsanto_2018,Schmidt_2020}, defined per object as the value of the predicted cumulative distribution function at the true redshift, $\mathrm{PIT} = \mathrm{CDF}(\zspec)$. For perfectly calibrated PDFs the PIT values are uniformly distributed; convex (concave) histograms indicate over- (under-) confident uncertainties, while slanted histograms indicate systematic bias. We report PIT histograms for all uncertainty-aware models on all four test subsets.

To quantify the agreement between the photometric and spectroscopic redshift distributions, we compute the mean squared error:
\begin{equation}
\label{eqn:mse_star}
    \msestar = \frac{10^3}{N}(\Szspec - \SPzphoto)^{\mathrm{T}}(\Szspec - \SPzphoto),
\end{equation}
where $N$ is the number of redshift bins, and $10^3$ is a normalizing factor. Additionally, we report the bias of its first and second moments in bins of redshift (width $\Delta z = 1$). In each bin we compare the mean $\langle z \rangle$ and standard deviation $\sigma$ of the stacked predicted distribution $\SPzphoto$ with those of the spectroscopic distribution $\Szspec$, and quote $\Delta \langle z \rangle$ and $\Delta \sigma$. We report these as a function of redshift. These two moments are the quantities to which two-point cosmological statistics (angular clustering, lensing cross-correlations) are primarily sensitive \citep{Ma_2006}; we do not treat them as a complete description of the $n(z)$ shape, which for quasars is frequently multimodal, nor do we attempt a tomographic assessment, which would depend on the selection binning of a specific cosmological analysis.

\subsection{Unsupervised learning}

In colour space, a phenomenon known as photo-z degeneracy may occur, whereby two or more distinct redshift values correspond to the same set of feature values. In such cases, uncertainty-aware models may produce bimodal or multimodal predictions, resulting in multiple local maxima in the estimated photo-z distribution. To identify these degeneracies, we perform an analysis based on t-distributed stochastic neighbor embedding \citep[t-SNE,][]{vanderMaaten_2008} and hierarchical density based clustering \citep[HDBSCAN,][]{McInnes_2017}, applied to a feature space built from the PDFs of photo-$z$ uncertainties.

We sample each $\Pzphoto$ at 100 evenly spaced redshift values between $z=0.05$ and $z=5.0$ and treat these as a feature vector. We embed this space into two dimensions with t-SNE, a non-linear dimensionality reduction that preserves local structure, using the correlation distance as the metric so that objects are grouped by the shape of their $\Pzphoto$ rather than its absolute values, with a perplexity of 40 and a learning rate of 200.

We subsequently apply the HDBSCAN to the embedded representations, with the minimum cluster size set to 50, and 5 samples required to form a core point. HDBSCAN is a density-based clustering tool that identifies groups of points closely packed together, while marking points in sparse regions as noise. Unlike K-Means, it automatically determines the number of clusters based on the data density structure. For each identified cluster, we construct a total redshift distribution histogram to examine its characteristic shape.

\section{Results}

We first evaluate the per-object PDFs through their negative log-likelihood (Sect.~\ref{sec:nll}) and their calibration (Sect.~\ref{sec:pit}), then compare the point estimates against the ANN (Sect.~\ref{sec:point}), then assess the reconstructed redshift distribution and the bias of its binned moments (Sect.~\ref{sec:nz}), next contrast these with the SOM estimate (Sect.~\ref{sec:som}), and finally examine the degeneracy structure of the predictions through unsupervised clustering (Sect.~\ref{sec:degeneracies}).

\subsection{Per-object PDF likelihood}
\label{sec:nll}

\begin{table}[hbt!]
\caption{Negative log-likelihood (NLL) of the predicted redshift PDFs for the MDN with one to six mixture components and the BNN with one to five components, evaluated on the four test subsets (Sect.~\ref{sec:testsets}). Each value is the mean over ten training realizations and the quoted uncertainty is their dispersion. The smaller the better; the smallest value in each column is highlighted in bold.}
\label{tab:nll}
\centering
\resizebox{\columnwidth}{!}{%
\begin{tabular}{lcccc}
\toprule
& \multicolumn{2}{c}{Complete features} & \multicolumn{2}{c}{Missing features} \\
\cmidrule(lr){2-3} \cmidrule(lr){4-5}
Model & Random test & Faint test & Random test & Faint test \\
\midrule
MDN 1 & $-0.17\pm0.02$ & $0.56\pm0.06$ & $0.53\pm0.17$ & $1.03\pm0.12$ \\
MDN 2 & \bfs $-0.66\pm0.01$ & $0.46\pm0.08$ & $0.34\pm0.18$ & $1.01\pm0.14$ \\
MDN 3 & $-0.66\pm0.03$ & $0.37\pm0.08$ & $0.47\pm0.32$ & $0.91\pm0.11$ \\
MDN 4 & $-0.66\pm0.03$ & $0.28\pm0.06$ & $0.39\pm0.25$ & $0.85\pm0.08$ \\
MDN 5 & $-0.66\pm0.03$ & $0.26\pm0.05$ & $0.38\pm0.24$ & $0.80\pm0.06$ \\
MDN 6 & $-0.65\pm0.03$ & $0.25\pm0.06$ & $0.57\pm0.38$ & $0.81\pm0.06$ \\
\midrule
BNN 1 & $-0.27\pm0.02$ & $0.51\pm0.03$ & $0.44\pm0.06$ & $1.50\pm0.21$ \\
BNN 2 & $-0.61\pm0.01$ & $0.19\pm0.03$ & $0.13\pm0.07$ & $0.83\pm0.06$ \\
BNN 3 & $-0.61\pm0.01$ & \bfs $0.13\pm0.02$ & \bfs $0.03\pm0.07$ & $0.71\pm0.06$ \\
BNN 4 & $-0.61\pm0.01$ & \bfs $0.13\pm0.02$ & $0.05\pm0.09$ & \bfs $0.70\pm0.03$ \\
BNN 5 & $-0.58\pm0.02$ & $0.17\pm0.04$ & $0.13\pm0.08$ & $0.82\pm0.10$ \\
\bottomrule
\end{tabular}%
}
\end{table}

Table~\ref{tab:nll} reports the NLL of the MDN, with one to six mixture components, and of the BNN, with one to five, on the four test subsets. Adding a second mixture component sharply reduces the NLL -- from $-0.17$ to $-0.66$ on the random test with complete features -- confirming that a single Gaussian cannot capture the multimodal quasar photo-$z$. Beyond two components the improvement saturates, with the MDN reaching its optimum at five components (its lowest NLL in the faint-extrapolation tests) and the BNN at three. Within each family the ranking of the components is preserved across the four subsets -- the ranking between the two families changes, as discussed below -- while the absolute NLL rises as extrapolation and missing bands are introduced.

Comparing the two best models in Table~\ref{tab:nll}, the five-component MDN and the three-component BNN, the BNN is worse by $0.05$ in NLL on the bright random test with complete features but better in every out-of-distribution regime: by $0.13$ under faint-magnitude extrapolation and by $0.09$ under simultaneous extrapolation and missing bands, with a nominal $0.35$ gain on the random test with missing features that, resting on only 261 objects, carries a correspondingly large scatter. The dispersion across the ten training realizations (Table~\ref{tab:nll}) sets the scale against which these differences should be read: the faint-test gains exceed it and are robust. The BNN therefore provides more realistic likelihoods precisely in the regimes that dominate a magnitude-limited quasar sample, at a modest cost on the brightest, best-covered data.

% -- Table 2 (tab:best) commented out to save space; the three carried-forward
% -- models are now identified in the text below. Kept here for reference.
% \begin{table*}[hbt!]
% \caption{Negative log-likelihood (NLL) of the three models carried forward into all subsequent analyses -- the single-component MDN, the five-component MDN, and the three-component BNN -- on the four test subsets (as in Table~\ref{tab:nll}). The lowest value in each column is highlighted in bold.}
% \label{tab:best}
% \centering
% \begin{tabular}{llcccc}
% \toprule
% & & \multicolumn{2}{c}{complete features} & \multicolumn{2}{c}{missing features} \\
% \cmidrule(lr){3-4} \cmidrule(lr){5-6}
% network & \#GMM & random test & faint test & random test & faint test \\
% \midrule
% MDN & 1 & -0.20 & 0.56 & 0.80 & 0.97 \\
% MDN & 5 & \bfs -0.70 & 0.24 & 0.07 & 0.83 \\
% BNN & 3 & -0.62 & \bfs 0.10 & \bfs -0.01 & \bfs 0.65 \\
% \bottomrule
% \end{tabular}
% \end{table*}

We therefore carry three models forward into all subsequent analyses -- the single-component MDN (the single-Gaussian baseline), the five-component MDN (the best multi-component MDN), and the three-component BNN (the best BNN). These are used for the calibration, the reconstructed $n(z)$ and its binned moments, and the comparison with the SOM estimate; the five-component MDN is additionally used for the degeneracy analysis. For the best single models, MDN~5 and BNN~3, and the four test subsets (random/complete, faint/complete, random/missing, faint/missing), the NLL is ($-0.70$, $0.24$, $0.07$, $0.83$) and ($-0.62$, $0.10$, $-0.01$, $0.65$), respectively.

\subsection{Per-object PDF calibration}
\label{sec:pit}

\begin{figure}[ht!]
    \centering
    \includegraphics[width=\hsize]{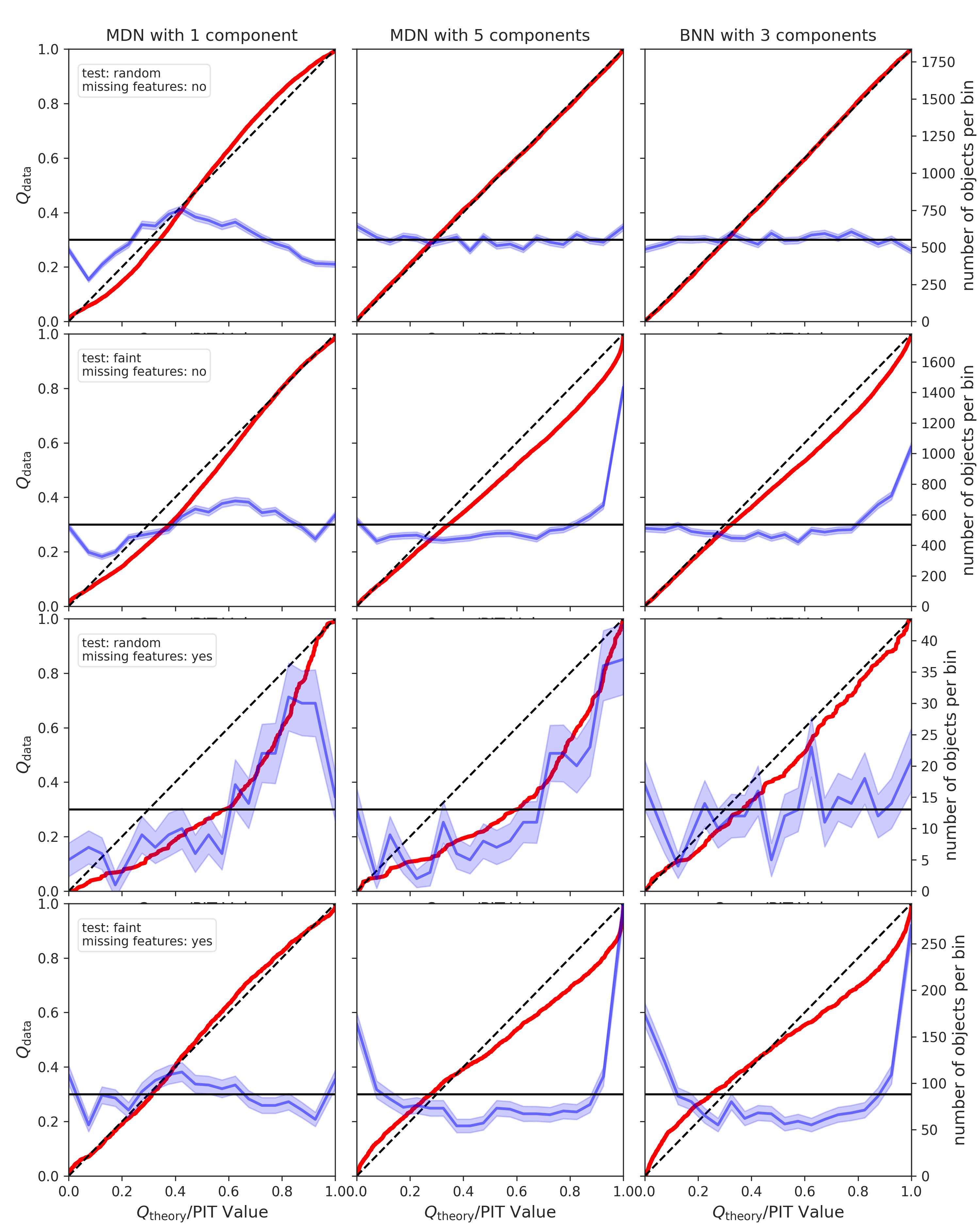}
    \caption{Probability integral transform (PIT) histograms of the predicted per-object redshift PDFs for the uncertainty-aware models. The blue and red curves are PIT densities and its quantile-quantile distributions, respectively. The horizontal line marks the uniform distribution expected for perfectly calibrated PDFs. We present MDN with one and five components, and a BNN with three components, shown for the four test subsets (Sect.~\ref{sec:testsets}).}
    \label{fig:pit}
\end{figure}

Figure~\ref{fig:pit} shows the PIT histograms. On the random test with complete features the five-component MDN and the three-component BNN are close to uniform, i.e. well calibrated, whereas the single-component MDN is visibly miscalibrated, its PIT peaked away from uniformity. Under faint-magnitude extrapolation a systematic departure appears for all models: the BNN remains the best calibrated, while the five-component MDN develops an over-confident high-redshift tail, seen as a pile-up of PIT values near unity. In the most difficult, faint-plus-missing regime this ordering reverses -- the single-component MDN yields the flattest PIT, while both the five-component MDN and the BNN become over-confident, their PIT piling up at $0$ and $1$ -- the signature of PDFs that are too narrow for their actual errors. The single Gaussian, unable by construction to place sharp, over-committed peaks, is therefore the most robustly calibrated model under the strongest distribution shift, consistent with its behaviour for the reconstructed $n(z)$ (Sect.~\ref{sec:nz}). The random test with missing features contains too few objects for its PIT to be informative.

\subsection{Point-estimate accuracy}
\label{sec:point}

\begin{table}[ht!]
\caption{Point-estimate quality of the photometric redshifts (mean of the predicted PDF) for the four test subsets (Sect.~\ref{sec:testsets}). We report the bias $\langle\delta z\rangle$ and scatter $\sigma(\delta z)$ of $\delta z = (\zphoto-\zspec)/(1+\zspec)$, and the outlier fraction $\eta$ (per cent of objects with $|\delta z|>0.15$). Each value is followed by its dispersion across the ten training realizations. The random test with missing features contains only 261 objects and is therefore statistically uncertain.}
\label{tab:point_estimates}
\centering
\resizebox{\columnwidth}{!}{%
\begin{tabular}{llccc}
\toprule
Test & Model & $\langle\delta z\rangle$ & $\sigma(\delta z)$ & $\eta$ [\%] \\
\midrule
\multirow{5}{*}{\shortstack[l]{random\\ complete}}
 & ANN   & $0.015\pm0.005$  & $0.133\pm0.003$ & $10.2\pm0.1$ \\
 & MDN 1 & $0.014\pm0.006$  & $0.128\pm0.003$ & $11.7\pm0.5$ \\
 & MDN 2 & $0.009\pm0.004$  & $0.121\pm0.002$ & $9.6\pm0.2$ \\
 & MDN 5 & $0.007\pm0.002$  & $0.121\pm0.002$ & $9.5\pm0.1$ \\
 & BNN 3 & $0.014\pm0.001$  & $0.124\pm0.002$ & $9.4\pm0.1$ \\
\midrule
\multirow{5}{*}{\shortstack[l]{faint\\ complete}}
 & ANN   & $-0.004\pm0.011$ & $0.157\pm0.003$ & $16.1\pm0.7$ \\
 & MDN 1 & $0.009\pm0.010$  & $0.153\pm0.003$ & $14.9\pm0.6$ \\
 & MDN 2 & $-0.012\pm0.006$ & $0.148\pm0.001$ & $14.9\pm0.7$ \\
 & MDN 5 & $-0.013\pm0.004$ & $0.147\pm0.001$ & $14.8\pm0.3$ \\
 & BNN 3 & $-0.007\pm0.005$ & $0.148\pm0.001$ & $14.1\pm0.3$ \\
\midrule
\multirow{5}{*}{\shortstack[l]{random\\ missing}}
 & ANN   & $0.006\pm0.012$  & $0.145\pm0.012$ & $9.2\pm1.3$ \\
 & MDN 1 & $0.003\pm0.027$  & $0.150\pm0.010$ & $13.9\pm1.8$ \\
 & MDN 2 & $-0.022\pm0.009$ & $0.135\pm0.011$ & $9.4\pm0.7$ \\
 & MDN 5 & $-0.032\pm0.010$ & $0.132\pm0.007$ & $9.3\pm1.0$ \\
 & BNN 3 & $0.005\pm0.009$  & $0.169\pm0.011$ & $11.6\pm0.9$ \\
\midrule
\multirow{5}{*}{\shortstack[l]{faint\\ missing}}
 & ANN   & $0.007\pm0.015$  & $0.198\pm0.004$ & $24.3\pm1.4$ \\
 & MDN 1 & $0.040\pm0.019$  & $0.200\pm0.010$ & $24.2\pm2.1$ \\
 & MDN 2 & $0.000\pm0.009$  & $0.191\pm0.007$ & $21.3\pm0.8$ \\
 & MDN 5 & $-0.003\pm0.006$ & $0.190\pm0.004$ & $21.1\pm0.6$ \\
 & BNN 3 & $0.014\pm0.011$  & $0.197\pm0.007$ & $21.4\pm1.0$ \\
\bottomrule
\end{tabular}%
}
\end{table}

Table~\ref{tab:point_estimates} reports, for the four test subsets, the bias and scatter of $\delta z$ and the outlier fraction $\eta$ of the point estimates (the mean of the predicted PDF). On the well-covered random test, reducing the PDFs to point estimates still outperforms the dedicated ANN regressor: the scatter tightens from $0.133$ to $0.121$ and the outlier fraction falls from $10.2\%$ to $9.6\%$ for the multi-component models. The single-component MDN is a clear exception -- it has the highest outlier fraction of all, $11.7\%$, exceeding even the ANN -- because a single Gaussian centred on one branch of a degenerate solution scatters the competing redshift into catastrophic outliers. Adding a second component removes this excess (down to $9.6\%$), and the improvement then saturates, mirroring the two-component threshold seen in the NLL. The same pattern holds out of distribution: in the reliable faint-plus-missing regime the outlier fraction drops from $24.2\%$ for the single-component MDN to $\simeq 21\%$ for the multi-component models, and the absolute bias from $0.040$ to $\lesssim 0.003$. Because tomographic bins are assigned from point estimates, these fewer catastrophic outliers translate directly into cleaner bins; this advantage of multi-component modelling for binning is distinct from, and complementary to, the reconstruction of $n(z)$ discussed next; we return to it in Sect.~\ref{sec:discussion}.

\subsection{Accuracy of the reconstructed $n(z)$}
\label{sec:nz}

\begin{figure*}
    \centering
    \includegraphics[height=0.92\textheight]{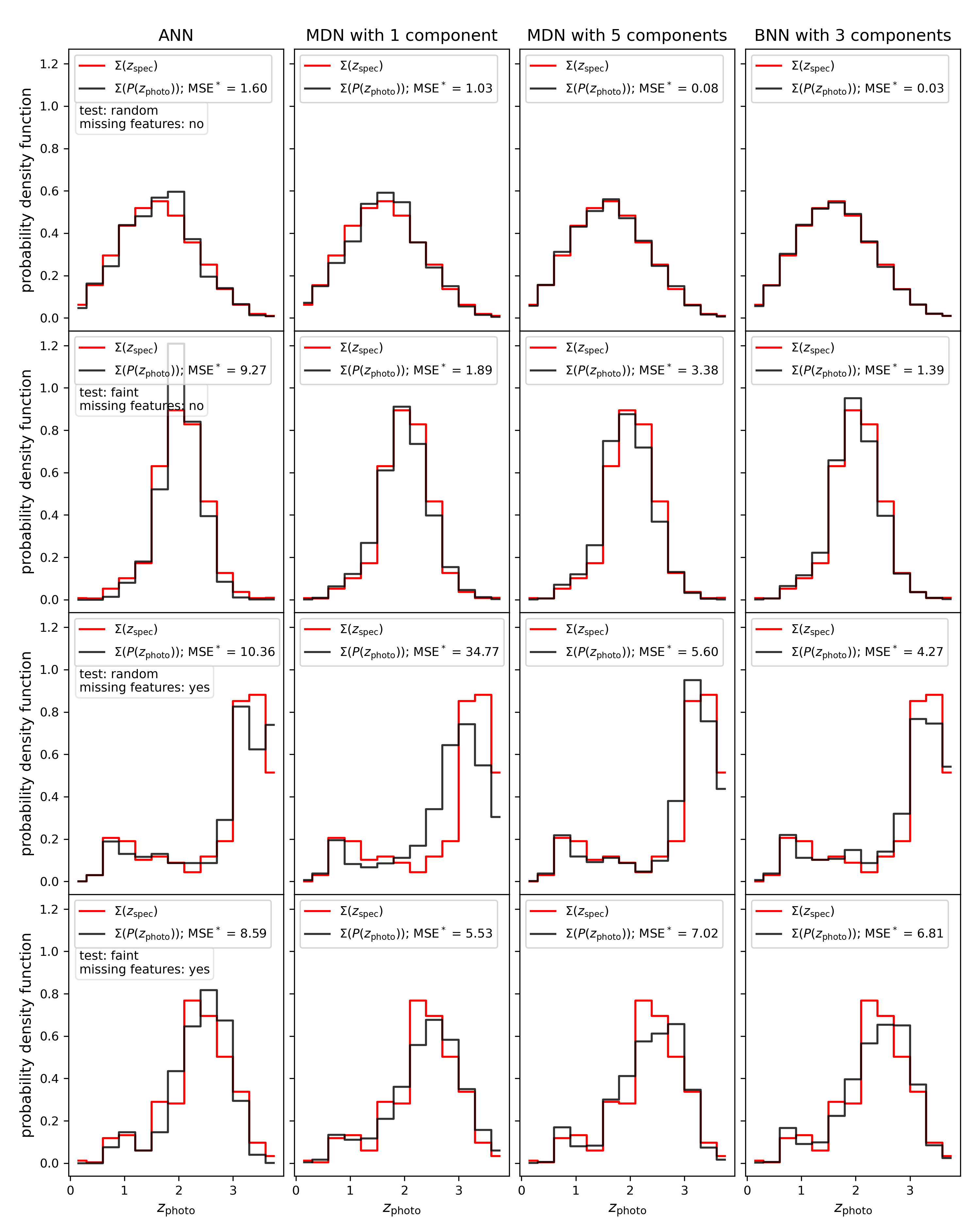}
    \caption{Spectroscopic (\textit{red}) and photo-$z$ (\textit{black}) redshift distributions for ANN (\textit{column 1}), MDN with 1 mixture component (\textit{column 2}), MDN with 5 mixture components (\textit{column 3}), and BNN with 3 mixture components (\textit{column 4}). Rows correspond to the four test subsets in the order defined in Sect.~\ref{sec:testsets}. We compute the mean squared error multiplied by a factor of $10^3$ ($\msestar$) for each model with respect to the spectroscopic distributions.}
    \label{fig:n_z}
\end{figure*}

Figure~\ref{fig:n_z} shows the reconstructed redshift distributions (black) against the spectroscopic truth (red). On the random test with complete features (top row) the uncertainty-aware models reproduce $n(z)$ almost perfectly: $\msestar=0.08$ for the five-component MDN and $0.03$ for the BNN. On the other hand, the ANN, at $\msestar=1.60$, is more than an order of magnitude worse and shows spurious step-like features characteristic of quasar point photo-$z$ estimates, which arise as broad emission lines redshift across successive broadband filters and drive a point estimator towards discrete redshift values. For the faint-magnitude extrapolation (second row) the BNN is the most accurate ($\msestar=1.39$); interestingly the five-component MDN ($3.38$) is here worse than the single-component MDN ($1.89$), because the additional components overshoot the sharp $z\simeq2.3$ peak. On the small random test with missing features (third row) the BNN again gives the lowest $\msestar$ ($4.27$) while the single-component MDN degrades strongly ($34.77$), although this subset of 261 objects does not give good statistics. In the combined faint-plus-missing regime (bottom row) all the models degrade ($\msestar$ between $5.53$ and $8.59$) and none reconstructs $n(z)$ reliably, the single-component MDN attaining the nominally lowest value ($5.53$). Thus, on well-covered data, the five-component MDN and the BNN reconstruct $n(z)$ best, whereas under the strongest shift the differences between models shrink and the simplest MDN is no longer disfavoured -- a pattern we quantify with the binned moments below and interpret in Sect.~\ref{sec:discussion}.

\begin{table*}[!hbt]
\caption{Bias of the first two moments of the reconstructed redshift distribution, $\Delta\langle z\rangle$ and $\Delta\sigma$, in redshift bins of width $\Delta z = 1$, for the ANN, the one- and five-component MDN, and the three-component BNN, on the four test subsets. The smallest absolute bias among the four models in each redshift bin is highlighted in bold.}
\label{tab:nz_moments}
\centering
\begin{tabular}{l l *{8}{c}}
\toprule
& & \multicolumn{2}{c}{$0 \leq z < 1$} & \multicolumn{2}{c}{$1 \leq z < 2$}
    & \multicolumn{2}{c}{$2 \leq z < 3$} & \multicolumn{2}{c}{$3 \leq z < 4$} \\
\cmidrule(lr){3-4} \cmidrule(lr){5-6} \cmidrule(lr){7-8} \cmidrule(lr){9-10}
Test & Model
& {$\Delta \langle z \rangle$} & {$\Delta \sigma$}
& {$\Delta \langle z \rangle$} & {$\Delta \sigma$}
& {$\Delta \langle z \rangle$} & {$\Delta \sigma$}
& {$\Delta \langle z \rangle$} & {$\Delta \sigma$} \\
\midrule
\multirow{4}{*}{\shortstack[l]{random\\ complete}}
       & ANN & -0.0247 & \bfs -0.0021 & -0.0303 & -0.0029 & -0.0032 & -0.0042 & 0.0154 & \bfs -0.0031 \\
& MDN 1 & 0.0202 & -0.0107 & -0.0275 & 0.0075 & \bfs -0.0013 & -0.0007 & 0.0332 & 0.0279 \\
& MDN 5 & \bfs -0.0130 & 0.0027 & \bfs -0.0015 & 0.0040 & -0.0071 & 0.0039 & 0.0157 & 0.0131 \\
& BNN 3 & -0.0154 & 0.0026 & -0.0039 & \bfs 0.0006 & \bfs -0.0013 & \bfs 0.0003 & \bfs -0.0091 & \bfs 0.0040 \\
\midrule
\multirow{4}{*}{\shortstack[l]{faint\\ complete}}
       & ANN & -0.1132 & 0.1435 & -0.0367 & \bfs 0.0007 & 0.0336 & 0.0123 & 0.1285 & 0.0926 \\
& MDN 1 & \bfs -0.0155 & \bfs 0.0627 & 0.0312 & -0.0203 & \bfs -0.0022 & -0.0184 & 0.0700 & 0.0907 \\
& MDN 5 & -0.0393 & 0.0811 & 0.0299 & -0.0066 & 0.0031 & -0.0106 & 0.1155 & 0.1192 \\
& BNN 3 & -0.0305 & 0.0674 & \bfs 0.0101 & -0.0095 &  0.0023 & \bfs -0.0054 & \bfs 0.0587 & \bfs 0.0744 \\
\midrule
\multirow{4}{*}{\shortstack[l]{random\\ missing}}
       & ANN & \bfs -0.0040 & 0.0182 & \bfs -0.0816 & \bfs -0.0122 & -0.0985 & 0.0539 & -0.0326 & -0.0075 \\
& MDN 1 & 0.0406 & -0.0467 & -0.1618 & -0.0202 & -0.0898 & 0.0639 & 0.0653 & 0.0043 \\
& MDN 5 &  0.0263 & \bfs -0.0166 &  -0.1068 &  -0.0168 & -0.1696 & 0.0670 & 0.0547 & 0.0077 \\
& BNN 3 & 0.0373 & -0.0415 & -0.1502 &-0.0262 & \bfs -0.0745 & \bfs 0.0294 & \bfs 0.0054 & \bfs -0.0010 \\
\midrule
\multirow{4}{*}{\shortstack[l]{faint\\ missing}}
       & ANN & -0.0860 & 0.1283 & -0.0496 & -0.0363 & \bfs -0.0213 & \bfs -0.0118 & 0.0899 & 0.0940 \\
& MDN 1 & 0.0265 & \bfs 0.0286 & \bfs 0.0068 & -0.0264 & -0.0317 & -0.0128 & -0.0412 & \bfs 0.0010 \\
& MDN 5 & -0.0040 & 0.0730 & -0.0537 & 0.0205 &  -0.0417 & -0.0213 & 0.0473 & 0.0483 \\
& BNN 3 & \bfs 0.0020 & 0.0703 & -0.0372 & \bfs -0.0055 & -0.0442 & -0.0167 & \bfs 0.0326 & 0.0342 \\
\bottomrule
\end{tabular}
\end{table*}

Table~\ref{tab:nz_moments} reports the bias of the first two moments of $n(z)$ per redshift bin. On the random test with complete features all four models -- including the point-estimate ANN and the single-component MDN -- have $|\Delta\langle z\rangle|\lesssim0.03$ in every bin, since averaging within a wide bin washes out per-object errors. For the faint-end extrapolation the biases grow, most strongly for the ANN (up to $|\Delta\langle z\rangle|\simeq0.13$ and $|\Delta\sigma|\simeq0.14$) and in the highest redshift bin, where the BNN shows the smallest bias ($\simeq0.06$) and the five-component MDN has the largest bias among the density models ($\simeq0.12$); notably the single-component MDN reaches a comparable or smaller moment bias than the five-component one. These low-order moments are, by construction, insensitive to the multimodal (degenerate) structure of the solutions (Sect.~\ref{sec:degeneracies}): a model can reproduce them well while misrepresenting the shape of the individual PDFs, which the point-estimate and degeneracy analyses probe. We report the moments per redshift bin because two-point statistics are primarily sensitive to them; we do not benchmark them against an external requirement, since none exists for the analyses in which photometric quasars are used.

\subsection{Comparison with the SOM estimate}
\label{sec:som}

\begin{figure}[ht!]
    \centering
    \includegraphics[width=\hsize]{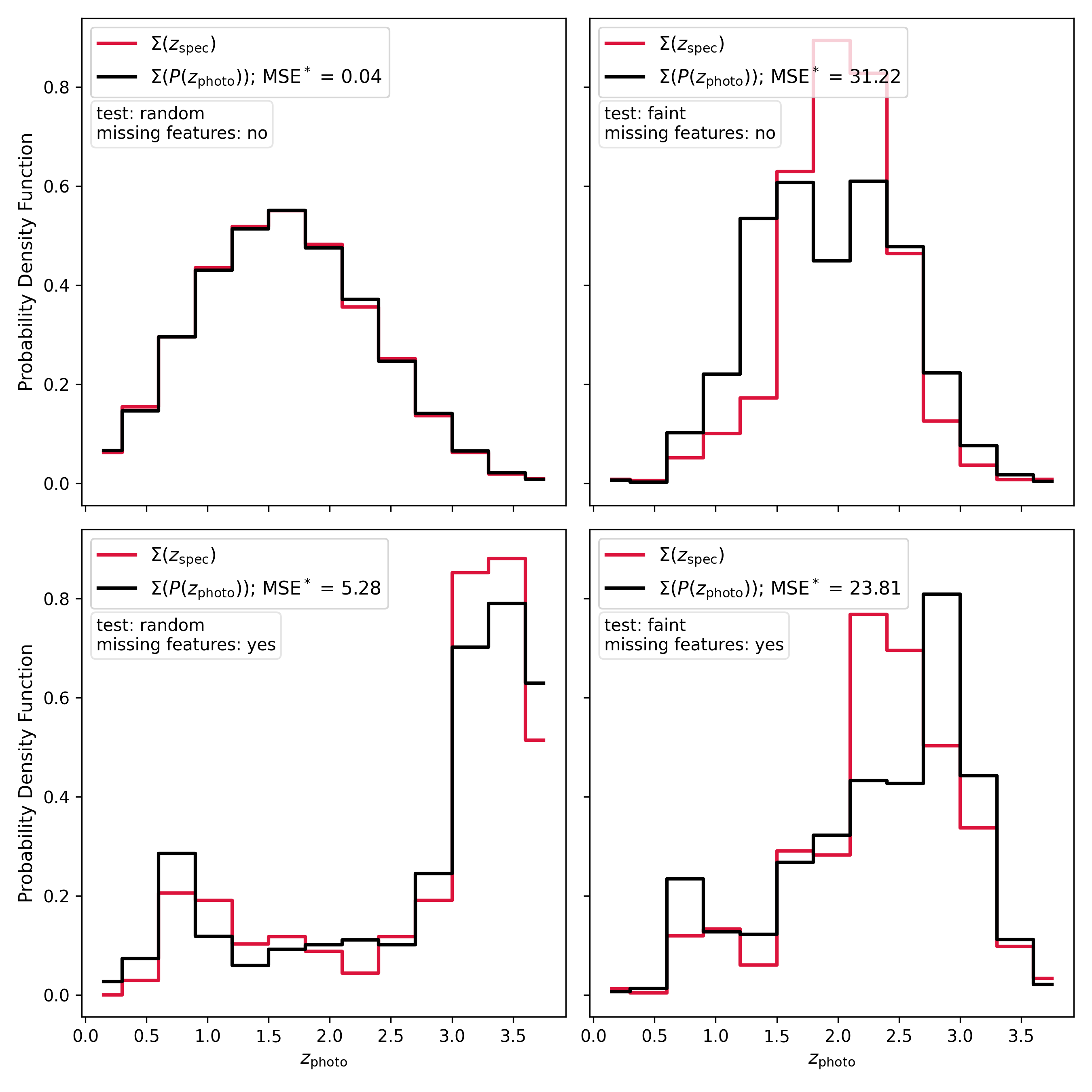}
    \caption{Redshift distributions $n(z)$ reconstructed by the SOM direct estimate (\textit{black}) compared with the spectroscopic distribution (\textit{red}), for the four test subsets (Sect.~\ref{sec:testsets}); each panel is labelled with the $\msestar$ of the SOM reconstruction. The corresponding $\msestar$ for the density models are listed in Table~\ref{tab:mse}.}
    \label{fig:som}
\end{figure}

\begin{table}[!hbt]
\caption{$\msestar$ ($\times10^3$; the smaller the better) quantifying the agreement between the stacked photometric and spectroscopic redshift distributions, for each model and test subset. The SOM row is an uncalibrated direct estimate (Sect.~\ref{sec:som_method}).}
\centering
\begin{tabular}{l *{4}{c}}
\toprule
& \multicolumn{2}{c}{Random} & \multicolumn{2}{c}{Faint} \\
\cmidrule(lr){2-3} \cmidrule(lr){4-5}
Model & Complete & Missing & Complete & Missing \\
\midrule
ANN           & 1.60      & 10.36     & 9.27      & 8.59 \\
MDN 1 & 1.03      & 34.77     & 1.89      & \bfs 5.53 \\
MDN 5 & 0.08      & 5.60      & 3.38      & 7.02 \\
BNN 3 & \bfs 0.03 & \bfs 4.27 & \bfs 1.39 & 6.81 \\
SOM           & 0.04      & 5.28      & 31.22     & 23.81 \\
\bottomrule
\end{tabular}
\label{tab:mse}
\end{table}

Figure~\ref{fig:som} shows the redshift distributions reconstructed by the SOM estimate, and Table~\ref{tab:mse} lists the corresponding $\msestar$ alongside those of the density models. We stress that the SOM is used here as a direct, non-parametric estimator of $n(z)$ \citep[as in e.g.][]{Geach_2012,Way_2012,CarrascoKind_2014}, not as a calibration scheme (Sect.~\ref{sec:som_method}). On the random test with complete features -- where the SOM cells are well populated by spectroscopic training objects -- the SOM reconstructs $n(z)$ essentially perfectly ($\msestar=0.04$), matching the best density models ($0.08$ for the five-component MDN, $0.03$ for the BNN), and it remains competitive also for the case of missing features ($5.28$). For the faint-magnitude extrapolation, however, it fails: $\msestar=31.22$ with complete features and $23.81$ with missing features, one to two orders of magnitude worse than the density models ($1.39$--$7.02$), smearing the sharp $z\simeq2$ peak into a broad plateau. The reason is that faint and incomplete objects are assigned to cells whose spectroscopic content is systematically brighter and at different redshift, so the empirical per-cell estimate is biased. The density models, by contrast, extrapolate beyond the spectroscopic coverage and propagate the associated uncertainty to their per-object PDFs, recovering the distribution more faithfully in exactly the regime that dominates a magnitude-limited quasar sample. We emphasize that this is the behaviour of the \emph{uncalibrated} SOM estimate; whether a dedicated redshift calibration would restore it is an open question that we take up in Sect.~\ref{sec:discussion}. This confirms the argument of Sect.~\ref{sec:intro}: for quasars, where discarding poorly covered regions of colour space is not an option, exploiting the faint population requires methods that extrapolate with per-object uncertainties.

\subsection{Photo-$z$ degeneracies}
\label{sec:degeneracies}

Fig. \ref{fig:t-sne} shows the t-SNE projections of photo-$z$ PDFs. We restrict this analysis to the held-out random test set with all features present and to the MDN with five components. The global structure resembles a spiral with several additional clusters. Objects for which the model predicts similar redshift values clearly cluster next to each other, forming a smooth gradient along the spiral: lower redshift values occupy the upper part of the spiral, while higher values appear toward the bottom. The distribution of standard deviation values remains relatively flat and shows comparable levels across most regions of the projection; the noticeable exceptions are the degenerate clusters 11 and 12, discussed below, where the uncertainties increase. Cluster $-1$ collects noise and outliers spread over a range of redshifts rather than a genuine degeneracy, and in our case corresponds to the smallest cluster.

To characterize the objects within each cluster in more detail, we construct photo-$z$ histograms of all objects in each cluster (Fig.~\ref{fig:clusters}). Clusters 11 and 12 exhibit clear signs of degeneracy: both the photometric and the spectroscopic histograms are bimodal, indicating that very similar colours are mapped to distinct redshifts. The degenerate redshift pairs are $(1.25, 1.85)$ for cluster~11 and $(0.85, 1.65)$ for cluster~12, and the two clusters contain $0.67\%$ and $0.44\%$ of the sample, respectively (about $1.1\%$ combined). We note that the full KiDS data, which is the final target for photo-z inference, contains a higher fraction of faint objects than the sample analysed here; the increased noise associated with these fainter sources may lead to a higher incidence of degenerate predictions, so this fraction should be read as a lower bound.

We examine whether clusters 11 and 12 could be characterized by specific features that would explain their degeneracy, but find no clear distinguishing properties. This suggests that, in the absence of sufficiently informative features, the model may express uncertainty through multi-peaked photo-$z$ estimates.

We stress that this analysis characterizes the degeneracy structure of the model's predictions on test data drawn from the same selection as the training sample, even if unseen by the model. We do not demonstrate that the same clusters or degeneracy patterns would be recovered on a target catalogue with a different selection function; applying the procedure directly to an inference catalogue, and verifying the stability of the recovered clusters, is a prerequisite to using it for sample cleaning. We further note that for two-point statistics, multimodal objects need not be removed at all, provided the redshift distribution is reconstructed accurately; removal is relevant primarily for use cases requiring reliable per-object redshifts, such as tomographic bin assignment or the selection of targets for follow-up.

\section{Discussion}
\label{sec:discussion}

Our results do not single out one best model to reconstruct the quasar redshift distribution; the ranking depends on the data regime and on the stage at which a tomographic analysis is considered. On the well-covered random test, the five-component MDN and the three-component BNN perform best in every aspect -- likelihood, calibration, point estimates and $n(z)$ -- and the SOM estimate is equally good. Out-of-distribution the picture splits. Considering NLL, the BNN is the best, reflecting sharper yet still reliable PDFs. For calibration and for the reconstructed $n(z)$ the advantage of the complex models over the single-component MDN shrinks and, in the most challenging faint-plus-missing regime, the situation reverses. The single Gaussian is then the best-calibrated model and reaches a comparable or smaller moment bias, because -- unable to place sharp, over-committed peaks -- it degrades slowly rather than becoming over-confident. At the binning stage, however, the ordering is unambiguous: the single-component MDN produces the most catastrophic outliers, more even than the ANN, and adding a second component removes this excess in every regime, including out of distribution.

Which of these aspects matters most depends on how the photometric redshifts are used. For two-point statistics with tomographic bins, the point estimates determine the bin assignment while the per-bin $n(z)$ impact the clustering kernels. Our results show larger model-to-model differences in the point estimates, and especially in their outlier fractions, than in the low-order moments of $n(z)$, where even the ANN performs reasonably well on covered data. A multi-component model is therefore clearly preferable for defining clean bins, whereas for the $n(z)$ itself the simplest model is competitive under strong extrapolation. Resolving this trade-off quantitatively -- how binning purity and $n(z)$ accuracy propagate into the angular power spectra $C_\ell$ -- requires calibrated, end-to-end pipelines that we defer to future work.

The improvement in the point estimates with mixture components (Sect.~\ref{sec:point}) is most naturally interpreted as a consequence of the greater sensitivity of a multi-component model to the detailed shape of each PDF. A better-modelled density yields a better central estimate and hence cleaner tomographic bins. Photo-$z$ degeneracies are the extreme limit of this, where a single Gaussian -- unable to place two modes -- fails outright and no intermediate solution is possible. In our data these two effects cannot be cleanly separated: the reduction in the outlier fraction with mixture components (Table~\ref{tab:point_estimates}) reflects better shape modelling in general, of which the bimodal clusters 11 and 12 are only the $\simeq1\%$ most severe cases. Degeneracies are expected to be considerably more frequent in the fainter, noisier inference catalogue, so isolating their specific impact on bin assignment -- with a dedicated, degeneracy-focused analysis and, ideally, probabilistic rather than fixed binning -- is a worthwhile extension that we leave to future work. We stress that, in any case, the low-order moments of $n(z)$ reveal none of these effects: a model can reproduce the distributions while misrepresenting the individual PDF shapes that drive the point estimates and the bin assignment.

None of the distributions presented here has been calibrated, and it is worth taking a closer look where the standard calibration methods would enter and what they can be expected to deliver for quasars. Two broad families exist. The first calibrates the redshift distribution of a sample or bin: `direct' k-nearest neighbour mapping of colours to redshift \citep[e.g.][]{Lima_2008,Cunha_2009,Hildebrandt_2017}, SOM reweighting \citep[e.g.][]{Buchs_2019,Wright_2020a,Wright_2020b,Hildebrandt_2021,Myles_2021,vandenBusch_2022,Wright_2025b,Yin_2025}, and clustering redshifts \citep[e.g.][]{Newman_2008,Menard_2013,Schmidt_2013,Gatti_2022}. The second calibrates the per-object PDFs directly, for example through a monotonic, PIT-based reshaping of each density \citep{Bordoloi_2010} or the more recent local, instance-wise calibration of conditional density estimates \citep{Dey_2022}. We note that a SOM that groups objects by their colours and maps them to the spectroscopic redshifts of each cell is itself an empirical, cell-based photometric-redshift estimator: its per-cell $p(z)$ is a per-object PDF \citep[e.g.][]{Geach_2012,Way_2012,CarrascoKind_2014}. Therefore, describing SOM reweighting as independent of the photo-$z$ algorithm is only partly accurate: it substitutes a low-capacity empirical estimator for the parametric one. Moreover, as our SOM comparison shows, it inherits the same distribution-shift failure where the calibrating spectroscopy is sparse. Clustering redshifts, which infer $n(z)$ from spatial cross-correlations and do not use colours, are immune to such issues; they however suffer from sparse sampling (especially for quasars) and the unknown galaxy bias redshift- and scale-dependence of the inference set \citep[e.g.][]{Matthews_2010,vandenBusch_2020,Gatti_2022,Newman_2022}.

Crucially, calibration is not guaranteed to improve the photo-z-based redshift distributions for quasars. All the colour-based methods above -- the direct and SOM reweighting schemes of the first family, as well as the per-object calibration of the second -- rest on the assumption that the colour--redshift relation is the same in the spectroscopic and photometric populations (covariate shift) and that the two overlap in colour space (the positivity, or common-support, condition; e.g. \citealp{Lima_2008,Autenrieth_2024}). For quasars both are fragile: the spectroscopic and photometric samples are selected very differently, and the faint photometric population extends into colour regions with little or no spectroscopic coverage, so the importance weights are ill-defined and the invariance of $p(z\,|\,\mathrm{colour})$ is not assured. Clustering redshifts are the exception: because they never use colours, they require neither covariate shift nor common support in colour space. They are, however, not assumption-free either -- the spectroscopic reference sample must cover the full redshift range of the target at sufficient density, and the redshift evolution of the galaxy bias of both samples has to be modelled. The reference selection function thus still propagates into the inferred $n(z)$, but through its clustering amplitude and redshift coverage, which are measurable from the reference sample itself, rather than through an unverifiable assumption about colour regions the spectroscopy does not reach. Methods such as StratLearn \citep{Autenrieth_2024} are demonstrably robust to covariate shift, but only for as long as these two conditions hold; whether they would help for faint quasars, where neither is assured, is an open question. We therefore regard our uncalibrated distributions not as a final product ready for cosmological analyses, but as a usable and well-characterized baseline, with the biases under extrapolation quantified in Table~\ref{tab:nz_moments}. Since the assumptions behind the standard calibration schemes are precisely the ones that break down in the faint regime dominating a magnitude-limited quasar sample, applying such a scheme on top of our estimates cannot be assumed to improve them: any future calibration study should demonstrate that it does. Our distributions are, moreover, already applicable to clustering and cross-correlation analyses in which $n(z)$ is marginalized over.

This calibration uncertainty has a direct consequence even in the well-covered regime. The single-component MDN is then not calibrated at all (Fig.~\ref{fig:pit}), so it could be used only after an external calibration, whereas the five-component MDN and the BNN are already well-calibrated on the covered test. If, as argued above, even the covered-regime calibration is itself uncertain for quasars because it does not model the DESI selection function, then a golden sample defined by trimming a photometric catalogue to the well-populated SOM cells -- that is, to in-distribution data -- is one on which only a multi-component MDN or a BNN can be trusted without a separate, and possibly unreliable, calibration step. Intrinsic calibration is therefore a decisive advantage precisely where one might have expected the simplest model to suffice.

Finally, our results should be interpreted within the scope of the experiment. Because the training and test sets are drawn from the same KiDS$\times$DESI quasar overlap, the absolute accuracies are an upper bound: they will not transfer unchanged to an inference catalogue with a different selection function, where the training--inference mismatch will dominate. The transferable results are the relative comparison of the uncertainty frameworks, the identification of the regimes in which each is preferable, and the various metrics and diagnostics used here -- NLL, PIT, the reconstructed $n(z)$ and its moments, the SOM comparison, and the degeneracy clustering. A quantitative, calibrated comparison of the full space of pipelines sketched in Sect.~\ref{sec:intro} is left to future work; the models characterized here provide its starting point, since the calibration would be applied directly to them and to the SOM, and evaluated with the same diagnostics.

\section{Conclusions}

In this work, we perform a comprehensive comparison of photometric redshift (photo-$z$) uncertainty estimation methods using machine learning techniques applied to quasars. We also assess the impact of photo-$z$ uncertainty on the reconstruction of the underlying redshift distribution. We provide a quantitative comparison and practical guidelines for addressing photo-$z$ prediction across a range of data quality scenarios and domain shifts. Our analysis is based on photometric observations from the KiDS combined with spectroscopically confirmed quasars from DESI, enabling a controlled evaluation under well-defined training conditions.

We demonstrate that an accurate reconstruction of the true redshift distribution is fundamentally unattainable without a robust uncertainty estimation framework. In particular, Mixture Density Networks (MDN) with at least two components are essential to capture the intrinsic multi-modality of photo-$z$ solutions. Within this family the negative log-likelihood is minimized by five components; however, no single model is the best across all regimes. The five-component MDN and the three-component BNN perform best on well-covered data and for point estimates, whereas at the faint-end extrapolation the single-component MDN becomes the best-calibrated model and remains competitive on the reconstructed $n(z)$.

Bayesian neural networks (BNNs) exhibit clear advantages in extrapolative regimes, where inference data extend beyond the magnitude range covered by the training set. In such cases, BNNs improve the negative log-likelihood by up to $0.18$ over the best MDN, indicating a more realistic uncertainty quantification, at the cost of a modest $0.08$ increase in NLL for well-covered, brighter sources. This trade-off highlights the context-dependent nature of Bayesian approaches in photo-$z$ estimation.

We find that redshift distribution reconstruction is highly accurate when inference data are complete and well represented in the training set. Moderate reconstruction errors emerge in scenarios involving either feature incompleteness or extrapolation beyond the training magnitude range, while the simultaneous presence of both effects leads to substantial degradation in reconstruction quality. These results emphasize the compounded impact of data incompleteness and domain shift on cosmological inference.

Furthermore, our unsupervised clustering analysis reveals two distinct clusters associated with degeneracies between two competing redshift solutions, at redshift pairs of $(1.25, 1.85)$ and $(0.85, 1.65)$. Unsupervised clustering of photo-$z$ predictions is thus an efficient diagnostic for such degeneracies, complementing traditional scalar metrics, and the degeneracies themselves may lead to incorrect bin assignment in tomographic analyses. Applying this diagnostic to sample cleaning would require verification on the target catalogue itself. Developing a method to resolve this degeneracy could provide the most effective pathway toward improving quasar photo-$z$ estimation.

Two further extensions concern the treatment of incomplete photometry. First, our experiments treat all objects with at least one missing magnitude equally; splitting this sample according to which specific band is missing would isolate the per-band impact on photo-$z$ quality, and our tests indicate this would be an informative refinement. Second, the origin of a missing measurement -- non-detection versus non-observation -- could not be established from the catalogue used here, but where ancillary survey metadata make this distinction possible, it could serve as an additional input feature for both quasar classification and photo-$z$ estimation. We leave both directions for future work.

It is important to note that this study is based on training data drawn from the overlap between the KiDS and DESI surveys, and therefore does not explicitly probe the additional effects of training–inference domain mismatch. As a result, our findings should be interpreted as an upper bound on achievable photo-$z$ performance. Future studies could investigate the impact of removing objects with the largest photo-$z$ uncertainties on the reconstructed redshift distribution, as well as explore approximate Bayesian techniques such as Monte Carlo dropout, which are widely adopted in deep learning applications.

The main methodological findings of this work are: i) the necessity of multi-component density outputs for capturing quasar photo-$z$ multimodality; ii) the regime-dependent trade-off of Bayesian networks; and iii) the compounded failure under simultaneous extrapolation and missing data. We believe they are all relevant for forthcoming large-scale surveys such as \textit{Euclid} \citep{Euclid_2022}, the Legacy Survey of Space and Time \citep[LSST,][]{Ivezic_2019}, and the Nancy Grace Roman Space Telescope \citep{Green_2012}. As stressed in Sect.~\ref{sec:discussion}, the absolute performance figures reported here will not transfer directly to such surveys; the relative comparison of uncertainty frameworks is the transferable result. These surveys will rely on precise redshift distributions for clustering and cross-correlation studies, often where training data are incomplete or unrepresentative and dedicated calibration is required. In this sense the present work is a first step towards the full comparative analysis of calibration pipelines that KiDS quasar photo-$z$ will ultimately require. The comparisons presented here provide a practical framework for selecting and validating photo-$z$ uncertainty methods, and for removing photo-$z$ degeneracies, in these next-generation experiments. Our eventual aim is to know not only quasar redshifts but how reliable those estimates are, in the regime where spectroscopy is no longer available. This will allow us to probe the faintest quasars and turn their full population into reliable tracers of the cosmic web.

\begin{acknowledgements}

We would like to thank Hendrik Hildebrandt for valuable discussions and for providing an internal review of this work. We also thank the anonymous journal referee for the very useful comments and suggestions that allowed us to considerably extend and improve the original manuscript.

SJN and MB are supported by the Polish National Science Center through grant no. 2020/38/E/ST9/00395 and by the Polish Ministry of Science and Higher Education under the agreement no. 2026/WK/06.

Based on data obtained from the ESO Science Archive Facility with DOI: \url{https://doi.org/10.18727/archive/37}, and \url{https://doi.eso.org/10.18727/archive/59} and on data products produced by the KiDS consortium. The KiDS production team acknowledges support from: Deutsche Forschungsgemeinschaft, ERC, NOVA and NWO-M grants; Target; the University of Padova, and the University Federico II (Naples).

The following software was used: Python 3 \citep{VanRossum_2009}, Tensorflow \citep{Abadi_2015}, Keras \citep{Chollet_2015}, Scikit-learn \citep{Pedregosa_2011}, NumPy \citep{Harris_2020}, SciPy \citep{Virtanen_2020}, IPython \citep{Perez_2007}, Pandas \citep{McKinney_2010}, Matplotlib \citep{Hunter_2007}, seaborn \citep{Waskom_2021}, tqdm \citep{daCostaLuis_2019}.

\end{acknowledgements}

\bibliographystyle{aa}
\bibliography{qso_photo-z}

\appendix

\section{Photometric-redshift degeneracies: t-SNE map and per-cluster distributions}
\label{app:degeneracies}

\begin{figure*}[h!]
    \centering
    \includegraphics[width=\textwidth]{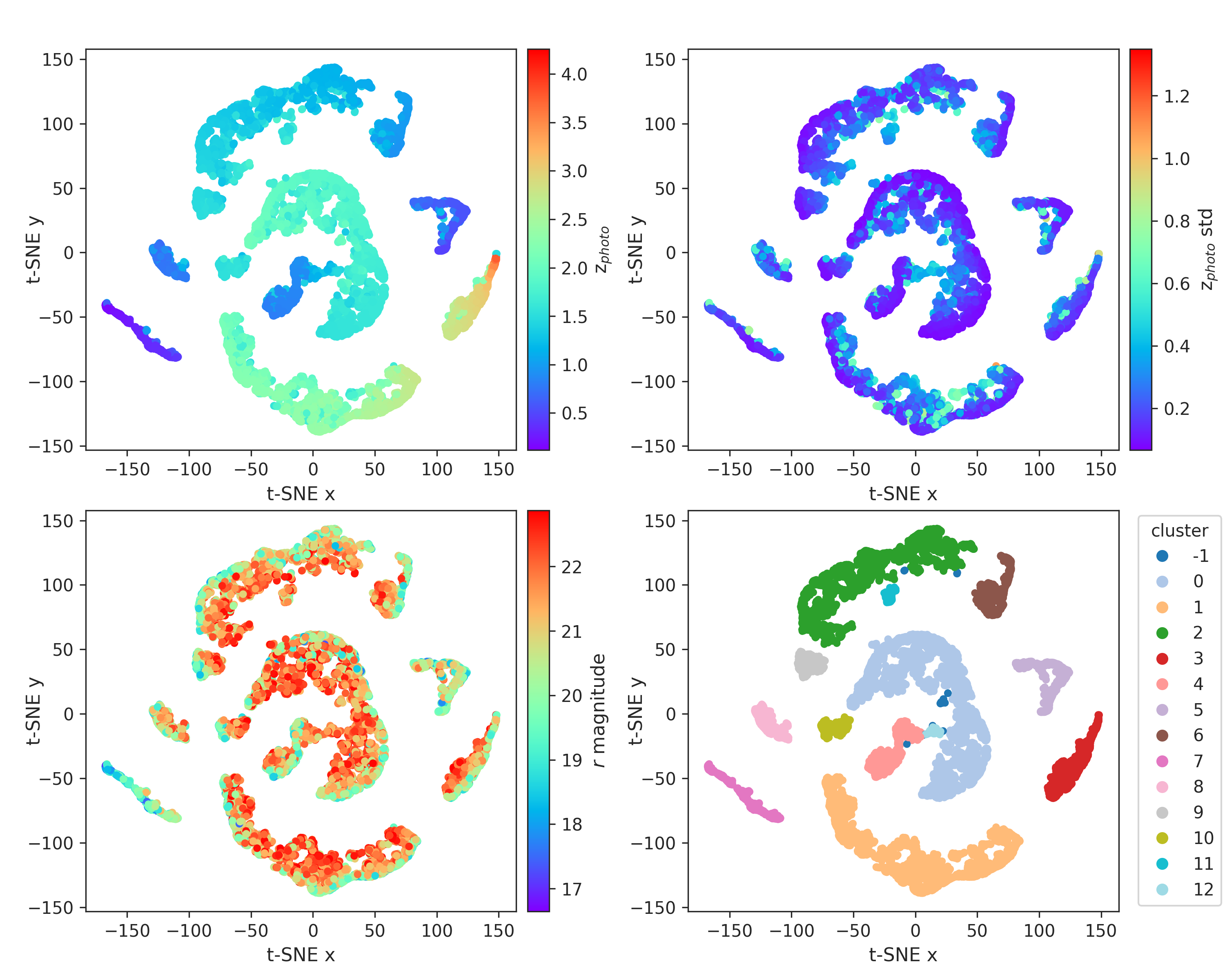}
    \caption{t-SNE projection of photo-$z$ estimates in the random test data with complete features. \textit{Top-left}: mean photo-$z$, \textit{top-right}: photo-$z$ standard deviation, \textit{bottom-left}: $r$ magnitude, \textit{bottom-right}: clusters. For each object, we use an array of equally spaced 100 probability density function (PDF) values calculated from model output.}
    \label{fig:t-sne}
\end{figure*}

\begin{figure*}
    \centering
    \includegraphics[height=0.90\textheight]{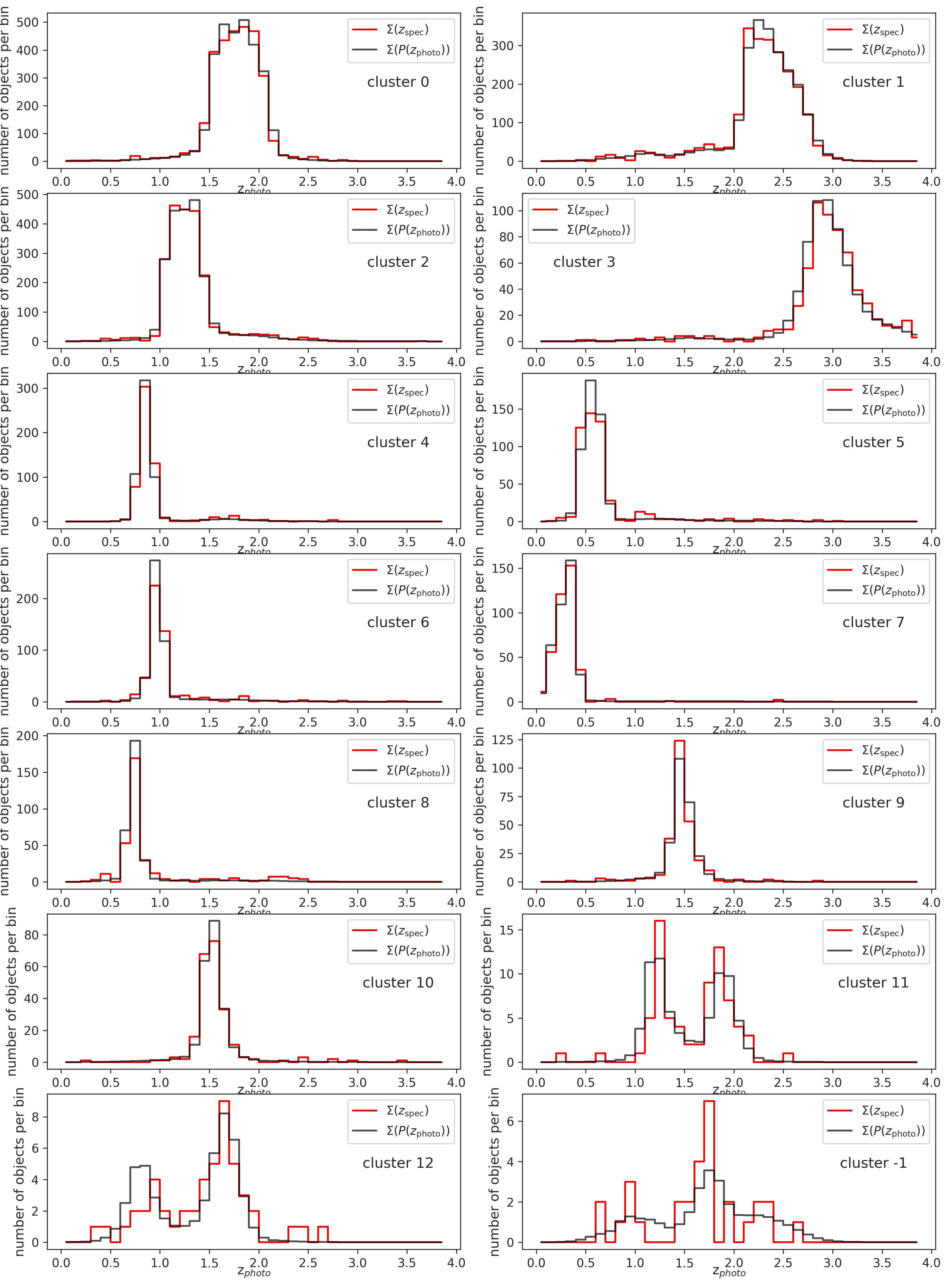}
    \caption{Spectroscopic and photometric redshift distributions for all clusters.}
    \label{fig:clusters}
\end{figure*}

This appendix collects the two figures supporting the photo-$z$ degeneracy analysis of Sect.~\ref{sec:degeneracies}: the t-SNE projection of the predicted PDFs (Fig.~\ref{fig:t-sne}) and the spectroscopic and photometric redshift distributions of the individual clusters (Fig.~\ref{fig:clusters}).

\end{document}